\documentclass{llncs}
\usepackage{makeidx}  
\usepackage{latexsym}
\usepackage{stmaryrd}
\usepackage{graphicx}
\usepackage{wrapfig}
\usepackage{paralist}
\usepackage{syntax}
\newcommand{\impl}{\Rightarrow}
\begin{document}

\title{\textsf{FMLtoHOL (version 1.0)}: Automating First-order Modal Logics with LEO-II and Friends\thanks{Supported by the German Research Foundation
         (BE2501/9-1 and KR858/9-1).}}
%
\titlerunning{Automating First-order Modal Logics with HOL-ATPs}  
%
\author{Christoph Benzm\"uller\inst{1} and Thomas Raths\inst{2}}
%
%
%
\institute{Freie Universit\"at Berlin, Germany
\and
University of Potsdam, Germany
}

\maketitle              

\begin{abstract}
  A converter from first-order modal logics to classical higher-order
  logic is presented. This tool enables the application of off-the-shelf
  higher-order theorem provers and model finders for
  reasoning within first-order modal logics. The tool supports
  logics K, K4, D, D4, T, S4, and S5 with respect to constant, varying
  and cumulative domain semantics.
\end{abstract}
\section{Introduction}
First-order modal logics (FMLs) \cite{FM98}
have many applications, e.g., in planning, natural language
processing, program verification, querying knowledge bases, and
modeling communication.  These applications motivate the use of {\em
  automated theorem proving\/} (ATP) systems for FMLs. 
Several new FML ATP systems, including two \textsf{FMLtoHOL}-based solutions, have recently been 
provided~\cite{Submitted}. 

This paper describes the \textsf{FMLtoHOL} tool, which automatically
converts problems in FML, formulated in the new
\texttt{qmf}-syntax~\cite{tr:RathsOtten12} (which extends the TPTP
\texttt{fol}-syntax \cite{Sut09} with operators \texttt{\#box} and
\texttt{\#dia}), into problems in classical higher-order logic (HOL),
formulated in \texttt{thf0}-syntax~\cite{SuBe10}. \textsf{FMLtoHOL} exploits and implements a semantic embedding of constant
domain FML in HOL~\cite{BP09,BP08}. Moreover, the tool
extends this embedding to varying and
cumulative domain semantics.

\textsf{FMLtoHOL} thus turns any \texttt{thf0}-compliant
HOL ATP system --- such as LEO-II\footnote{Cf.~{\tt www.leoprover.org} and {\tt www.ps.uni-saarland.de/\~{}cebrown/satallax/}} and Satallax$^3$ --- into a flexible ATP system for FML. At present \textsf{FMLtoHOL}
supports modal logics $L:=\{$K,K4,D,D4,T,S4,S5$\}$, all with respect to
constant, varying and cumulative domain semantics. 
Extending the tool to other normal FMLs 
and their combinations 
is straightforward.

In the remainder the language of FML is fixed as: $F,G ::= 
P(t_1,\ldots,t_n)
\mid \neg F \mid F \wedge G \mid
F \vee G \mid F \impl G \mid \Box F \mid
\Diamond F \mid \forall{x} F \mid \exists{x} F$.
The symbols $P$ are $n$-ary ($n \geq 0$) relation constants which are applied to terms
$t_1,\ldots,t_n$. The $t_i$ ($0\leq i\leq n$) are ordinary first-order terms
 and they may contain function symbols. 
The formula 
$(\forall x \Box fx) \impl \Box \forall x fx$ is used as an example,
it is referred to as \texttt{E1}. In constant domain (resp.~varying
domain) semantics \texttt{E1} is a theorem (resp.~countersatisfiable)
for logics $L$. In cumulative domain semantics
\texttt{E1} is a theorem
for S5 and countersatisfiable for the other logics in $L$.


%
\section{Theory of \textsf{FMLtoHOL}}
\textsf{FMLtoHOL} exploits the fact that Kripke structures can be
elegantly embedded in HOL~\cite{BP09,BP08}: FML
propositions $F$ are associated with HOL terms $F_\rho$ of predicate
type $\rho:=\iota\shortrightarrow o$. Type $o$
denotes the set of truth values and type $\iota$ is associated with the
domain  of possible worlds. Thus, the application $(F_\rho
w_\iota)$ corresponds to the evaluation of FML proposition $F$ in
world $w$. Consequently, validity is formalized as
$vld_{\rho\shortrightarrow o}=\lambda {F_{\rho}} \forall w_\iota F w$.
Classical connectives like $\neg$ and $\vee$ are simply lifted to type
$\rho$ as follows: $\neg_{\rho\shortrightarrow\rho} = \lambda
{F_{\rho}}\lambda {w_\iota} \neg F w$ and
$\vee_{\rho\shortrightarrow\rho\shortrightarrow\rho} = \lambda
{F_{\rho}} \lambda {G_{\rho}} \lambda {w_\iota} (F w \vee G w)$.
$\Box$ is modeled as $\Box_{\rho\shortrightarrow\rho} = \lambda
{F_{\rho}} \lambda {w_{\iota}} \forall {v_{\iota}} (\neg R w v \vee F
v)$, where constant symbol $R_{\iota\shortrightarrow\rho}$ denotes the
accessibility relation of the $\Box$ operator, which remains
unconstrained in logic K. Further logical connectives are defined as
usual: ${\wedge} = \lambda {F_{\rho}} \lambda {G_{\rho}} \neg (\neg F
\vee \neg G)$, $ {\Rightarrow} =\lambda {F_{\rho}} \lambda {G_{\rho}}
(\neg F \vee G)$, $\Diamond = \lambda {F_{\rho}} \neg \Box \neg F$.
To obtain e.g. modal logics D, T, S4, and S5, $R$ is axiomatized as
serial, reflexive, reflexive and transitive, and an equivalence
relation, respectively. Arbitrary normal modal logics extending K can
be axiomatized this way.

For individuals a further base type $\mu$ is reserved in
HOL. Universal quantification $\forall x F$ is introduced as syntactic
sugar for $\Pi\lambda {x} F$, where $\Pi$ is defined as follows:
$\Pi_{(\mu\shortrightarrow\rho)\shortrightarrow\rho} = \lambda
{H_{\mu\shortrightarrow\rho}} \lambda {w_\iota} \forall {x_\mu} H x
w$. For existential quantification, $\Sigma = \lambda
{H_{\mu\shortrightarrow\rho}} \neg \Pi \lambda x_\iota\neg H x$ is
introduced. $\exists x F$ is then syntactic sugar for $\Sigma\lambda
{x} F$.
$n$-ary relation symbols P, $n$-ary function symbols $f$ and
individual constants $c$ in FML obtain types
$\mu_1\shortrightarrow\ldots\shortrightarrow\mu_n\shortrightarrow\rho$,
$\mu_1\shortrightarrow\ldots\shortrightarrow\mu_n\shortrightarrow\mu_{n+1}$
(both with $\mu_i=\mu$ for $0\leq i \leq n+1$) and $\mu$,
respectively.  

For any FML formula $F$ holds:   $F$ is a valid in modal logic K  for constant
  domain semantics if and only if $vld\,F_\rho$ is valid in HOL
  for Henkin semantics. This correspondence provides the foundation for proof automation of FMLs with HOL-ATP systems. 
  The correspondence follows from~\cite{BP09}, where a more general result
is shown for FMLs with additional quantification over Boolean variables.

The above approach is adopted for varying domain semantics as follows:
\begin{inparaenum}
\item \label{a} $\Pi$ is now defined as  $\Pi = \lambda H_{\mu\shortrightarrow\rho} \lambda
  {w_\iota}  \forall {x_{\mu}}  \texttt{exInW} x w \Rightarrow H x  w$, where relation
  $\texttt{exInW}_{\mu\shortrightarrow\iota\shortrightarrow o}$ (for 'exists in
  world') relates individuals with worlds. 
\item \label{b} The non-emptiness axiom $\forall {w_\iota} \exists
  {x_\mu} \texttt{exInW} x w$  for these individual domains is added.
\item \label{c} For
  each individual constant symbol $c$  an axiom $\forall {w_\iota} \texttt{exInW} c w$ is postulated; these axioms enforce the designation of $c$ in the individual domain of each world $w$. Analogous designation axioms are required for function symbols.
\end{inparaenum}

For cumulative domain semantics the axiom $\forall
x_\mu\forall v_\iota\forall w_\iota \texttt{exInW} x v \wedge R v w \Rightarrow
\texttt{exInW} x w$ is additionally postulated. It states that the individual domains are increasing along  relation $R$.




\section{Implementation and Functionality of \textsf{FMLtoHOL}} 
\textsf{FMLtoHOL} is implemented as part of the \textsf{TPTP2X}
tool~\cite{Sut09}. \textsf{TPTP2X} is a multi-functional utility for
generating, transforming, and reformatting TPTP problem files. It is
written in Prolog and it can be easily modified and extended.

The tool is invoked as ``\texttt{tptp2X -f
  thf:<logic>:<domain>  <qmf-file>}'' where $\texttt{<logic>} \in\{$K,K4,D,D4,T,S4,S5$\}$ and $\texttt{<domain>} \in\{const,vary,cumul\}$.

To illustrate its use it is assumed that file \texttt{E1.qmf} contains \texttt{E1} in \texttt{qmf}-syntax:
{\footnotesize
\begin{verbatim}
qmf(con,conjecture,( 
    ( ! [X] : ( #box : ( f(X) ) ) ) => ( #box : ( ! [X] : ( f(X) ) ) ) )).
\end{verbatim}
} ``\texttt{tptp2X -f
  thf:d:const E1.qmf}'' generates the corresponding HOL problem file \texttt{E1.thf} in \texttt{thf}-syntax\footnote{Some explanations: \texttt{\^{}} is $\lambda$-abstraction and \texttt{@} an (explicit) application operator. \texttt{!}, \texttt{?}, \texttt{\~{}}, \texttt{|}, and  \texttt{=>}  encode universal and existential quantification, negation, disjunction and implication in HOL. \texttt{mu > \$i > \$o} encodes the HOL type $\mu\shortrightarrow\iota\shortrightarrow o$. 
\texttt{mimplies}, \texttt{mforall_ind}, and  \texttt{mbox_d} are embedded logical connectives as described in Sect.~2. Their denotation is fixed by adding definition axioms; see e.g.~\texttt{mforall_ind} below.} \cite{SuBe10}  for constant domain logic D:
{\footnotesize
\begin{verbatim}
%----Include axioms for modal logic D under constant domains
include('Axioms/LCL013^0.ax.const').
include('Axioms/LCL013^2.ax').
%------------------------------------------------------------------------
thf(f_type,type,( f: mu > $i > $o )).   % type declaration for constant f

thf(prove,conjecture,( mvalid @ 
    ( mimplies @ ( mforall_ind @ ^ [X: mu] : ( mbox_d @ ( f @ X ) ) ) 
               @ ( mbox_d @ ( mforall_ind @ ^ [X: mu] : ( f @ X ) ) ) ) )).
\end{verbatim}
} The included axiom files contain the definitions of the logical
connectives as outlined in Sect.~2. For example, the definition for \texttt{mforall_ind} (which realizes $\Pi$ for constant domain semantics) is given in \texttt{LCL013\^{}0.ax.const}:
{\footnotesize
\begin{verbatim}
thf(mforall_ind,definition,( mforall_ind = 
    ( ^ [Phi: mu > $i > $o, W: $i] : ! [X: mu] : ( Phi @ X @ W ) ) )).
\end{verbatim}
}
\noindent  \texttt{LCL013\^{}2.ax} contains the definition of the serial $\Box$ operator in logic D:
{\footnotesize
\begin{verbatim}
thf(mbox_d,definition,( mbox_d = 
    ( ^ [Phi: $i > $o,W: $i] :
      ! [V: $i] : ( ~ ( rel_d @ W @ V ) | ( Phi @ V ) ) ) )).

thf(a1,axiom,( mserial @ rel_d )).
\end{verbatim}
} 
\noindent Similar definitions are provided in the included axiom files
for the other logical connectives and for auxiliary terms like
\texttt{mserial}.  The HOL ATP systems LEO-II and Satallax when
applied to \texttt{E1.thf} find a proof within a few milliseconds.

When \textsf{FMLtoHOL} is called with option
``\texttt{-f thf:s5:vary}'' a modified file
\texttt{E1.thf} is created containing a conjecture identical to
above except that \texttt{mbox_d} is replaced by
\texttt{mbox_s5}. Moreover, \texttt{E1.thf} now includes different
axiom files \texttt{LCL013\^{}0.ax.vary} and \texttt{LCL013\^{}6.ax}. The former contains a modified definition of \texttt{mforall_ind}, which incorporates an explicit
'exists in world' condition:
%
{\footnotesize
\begin{verbatim}
thf(mforall_ind,definition,( mforall_ind = 
    ( ^ [Phi: mu > $i > $o,W: $i] :
      ! [X: mu] : ( ( exists_in_world @ X @ W ) => ( Phi @ X @ W ) ) ) )).

thf(nonempty_ax,axiom,(
    ! [V : $i] : ? [X : mu] : (exists_in_world @ X @ V))).
\end{verbatim}
}
\noindent The latter axiom specifies the domains of existing objects
as non-empty for all worlds worlds $V$. Axiom file
\texttt{LCL013\^{}6.ax} specifies \texttt{mbox_s5} as follows:
{\footnotesize
\begin{verbatim}
thf(mbox_s5,definition,( mbox_s5 = 
    ( ^ [Phi: $i > $o,W: $i] :
      ! [V: $i] : ( ~ ( rel_s5 @ W @ V ) | ( Phi @ V ) ) ) )).

thf(a1,axiom,( mreflexive @ rel_s5 )).
thf(a2,axiom,( mtransitive @ rel_s5 )).
thf(a3,axiom,( msymmetric @ rel_s5 )).
\end{verbatim}
}
For the modified problem Satallax finds a counter model within milliseconds.

\section{Discussion and Outlook}
The \texttt{FMLtoHOL} has been applied and evaluated in combination
with the HOL ATP systems Satallax and LEO-II; cf.~\cite{Submitted} for
details. In this case study the approach has also been compared with
other, heterogeneous FML ATP systems.
The \texttt{FMLtoHOL} based solution has the best coverage (and it can
easily be extended to other modal logics and their combinations) and it is second best in overall
performance behind the clausal connection prover \textsf{MleanCoP}\footnote{Cf.~the information at {\tt\,http://www.iltp.de/qmltp/systems.html}}.

Future work includes several optimizations of the tool,
extensions for multimodal logics (which it already partly supports),
and further case studies. These case studies should evaluate the tool also in combination with other \texttt{thf0}-compliant HOL provers and model finders as outlined in~\cite{SuBe10}: TPS, Isabelle, Refute and Nitpick.

A recent observation is that the HOL model finders Satallax, Refute
and Nitpick apparently work well for constant and varying domain
semantics but have problems to find counter models for
cumulative domain semantics.\\[-1.5em]
\paragraph{Acknowledgements:} We thank Jens Otten, Geoff Sutcliffe and Chad Brown for their support and their valuable input to this work.


\begin{thebibliography}{12}\small
\vspace*{-.5em}









\bibitem{Submitted} C. Benzm\"uller, T.~Raths and J.~Otten. {Implementing and
    Evaluating Provers for First-order Modal Logics}, Proceedings of
  ECAI'2012, to appear.

\bibitem{BP09} C. Benzm\"uller and L. C. Paulson.  Quantified Multimodal
  Logics in Simple Type Theory.  Logica Universalis, 2012. DOI:10.1007/s11787-012-0052-y

 \bibitem{BP08} C. Benzm\"uller and L. C. Paulson, Exploring Properties of Normal Multimodal Logics in Simple Type Theory with LEO-II. In Festschrift in Honor of Peter B. Andrews on His 70th Birthday, 2008. College Publications.















\bibitem{FM98} M. Fitting and R. L. Mendelsohn.
 {\em First-Order Modal Logic\/}. Kluwer, 1998.


  





















\bibitem{tr:RathsOtten12}
T.~Raths and J.~Otten.
The QMLTP Problem Library for First-order Modal Logics. The 6th International Joint Conference on Automated Reasoning (IJCAR), Manchester, UK, 2012.









\bibitem{SuBe10} G. Sutcliffe and C. Benzm\"uller. Automated Reasoning in Higher-Order Logic using the TPTP THF Infrastructure. Journal of Formalized Reasoning, 3(1):1-27, 2010.







\bibitem{Sut09}
G. Sutcliffe.
\newblock {The TPTP Problem Library and Associated Infrastructure:
                    The FOF and CNF Parts, v3.5.0}.
\newblock {\em Journal of Automated Reasoning\/}, 43(4):337--362, 2009.






\end{thebibliography}
\end{document}